\documentclass{aa}  

\usepackage{graphicx}
\usepackage{txfonts}
\usepackage[colorlinks]{hyperref}
\begin{document}

   \title{Can one-zone hadronuclear model explain the hard-TeV spectrum of BL Lac objects?}


   \author{Wei-Jian Li
          \inst{1}
          \and
          Rui Xue\inst{1}
          \and
          Guang-Bo Long\inst{2}
          \and
          Ze-Rui Wang\inst{3}
          \and
          Shigehiro Nagataki\inst{4,5}
          \and
          Da-Hai Yan\inst{6,7}
          \and 
          Jian-Cheng Wang\inst{6,7}
          }

   \institute{Department of Physics, Zhejiang Normal University, Jinhua 321004, People's Republic of China\\
              \email{ruixue@zjnu.edu.cn}
    \and
        School of Physics, Sun Yat-sen University, Guangzhou, GuangDong, People's Republic of China
         \and
             College of Physics and Electronic Engineering, Qilu Normal University, Jinan 250200, People's Republic of China
             \and
             Astrophysical Big Bang Laboratory (ABBL), RIKEN, Saitama 351-0198, Japan
             \and
             Interdisciplinary Theoretical \& Mathematical Science Program (iTHEMS), RIKEN, Saitama 351-0198, Japan
             \and
             Yunnan Observatories, Chinese Academy of Sciences, Kunming 650011, China
             \and
             Key Laboratory for the Structure and Evolution of Celestial Objects, Chinese Academy of Sciences, Kunming 650011, China
             }

   \date{Received xxx; accepted xxx}

 
  \abstract
   {The intrinsic TeV emission of some BL Lacs are characterized by a hard spectrum (the hard-TeV spectrum) after correcting for the extragalactic background light. The hard-TeV spectra pose a challenge to conventional one-zone models, including the leptonic model, the photohadronic model, the proton synchrotron model, etc.}
   {In this work, we study if the one-zone hadronuclear ($pp$) model can be used to interpret the hard-TeV spectra of BL Lacs without introducing extreme parameters.}
   {We give analytical calculations to study if there is a parameter space and the charge neutrality condition of jet can be satisfied when interpreting the hard-TeV spectra of BL Lacs without introducing a super-Eddington jet power.}
   {We find that in a sample of hard-TeV BL Lacs collected by \cite{2019ApJ...871...81X}, only the hard-TeV spectrum of 1ES 0229+200 could be explained by $\gamma$-ray from $\pi_0$ decay produced in the $pp$ interactions, but at the cost of setting a small radius of the radiation region that comparable to the Schwarzschild radius of the central black hole. Combining with previous studies of other one-zone models, we suggest that the hard-TeV spectra of BL Lacs cannot be explained by any one-zone models without introducing extreme parameters, and should originate from the multiple radiation regions.}
   {}


   \maketitle
%

\section{Introduction}

Blazars are a class of active galactic nuclei (AGNs) with their relativistic jets pointing to the observer \citep{1995PASP..107..803U}. Combining with multi-wavelength observations, it is found that the spectral energy distributions (SEDs) of blazars usually exhibit two characteristic bumps. It is generally accepted that the low-energy bump originates from the synchrotron radiation of primary relativistic electrons in the jet. While, the origin of the high-energy bump is under debate. In leptonic models, the high-energy bump is explained by inverse Compton (IC) radiation from relativistic electrons that up-scatter soft photons emitted by the same population of electrons \citep[synchrotron-self Compton, SSC;][]{1985ApJ...298..114M}, or soft photons from external photon fields (external Compton, EC) such as the accretion disk \citep{1993ApJ...416..458D}, the broad-line region \citep[BLR,][]{1994ApJ...421..153S} or the dusty torus \citep[DT,][]{2000ApJ...545..107B}. In hadronic models, the high-energy bump is supposed to be originated from proton synchrotron radiation \citep{2000NewA....5..377A}, emission from secondary particles generated in photohadronic processes ($p\gamma$) including photopion and Bethe-Heitler pair production processes \citep{2019ApJ...884L..17S}, and the internal $\gamma \gamma$ pair production, or cascade emission generated in the intergalactic space through $p\gamma$ interactions by ultra-high energy cosmic rays (UHECR) with energies up to $10^{19 - 20}$ eV beamed by the blazar jet \citep[e.g.,][]{2011APh....35..135E}.

The extragalactic TeV background is dominated by emission from blazars \citep{2016PhRvL.116o1105A}, most of which are BL Lacertae objects\footnote{http://tevcat.uchicago.edu/} \citep[BL Lacs;][]{1995PASP..107..803U}. After correcting for the extragalactic background light (EBL) absorption, the obtained intrinsic TeV emission of some BL Lacs often shows a hard spectrum, i.e., the photon index $\Gamma_{\rm TeV} < 2$ (hereafter the hard-TeV spectrum). However, some intrinsic TeV spectra are too hard to be interpreted by conventional radiation models. In the modeling of blazars' radiation, it is usually assumed that all the jet's non-thermal emission comes from a compact spherical region, named blob. Such a model is also known as the one-zone model, where the one-zone leptonic model is the most commonly used \citep[e.g.,][]{2014Natur.515..376G, 2020ApJS..248...27T, 2021MNRAS.506.5764D}. However, since the Klein-Nishina (KN) effect softens the IC emission in the TeV band naturally, the one-zone leptonic model cannot reproduce the hard-TeV spectrum unless setting an extremely high Doppler factor \citep[e.g.,][]{2012A&A...544A.142A}, which is in conflict with radio observations \citep{2009A&A...494..527H}, and/or introducing a very high value of the minimum electron Lorentz factor \citep[e.g.,][]{2006MNRAS.368L..52K}. In the modeling of blazars, the one-zone $p\gamma$ model is also widely applied. To explain the hard-TeV spectrum, an extremely high proton power is required since the $p\gamma$ interaction is very inefficient. Based on the fact that the interaction efficiency of the photopion process in the high-energy limit is about 1000 times smaller than the bump of $\gamma \gamma$ opacity, \cite{2019ApJ...871...81X} prove that the required minimum jet powers of a sample of hard-TeV BL Lacs exceed the corresponding Eddington luminosities of the supermassive black holes (SMBHs). For the emitting region with a strong magnetic field (10--100~G), the one-zone proton synchrotron model is also widely applied to explain the high-energy bump \citep{2013ApJ...768...54B}. However, the super-Eddington jet power is also needed \citep{2015MNRAS.450L..21Z} except for some few cases such as the fast variabilities \citep[$t_\nu$ $\leq10^3$s;][]{2016ApJ...825L..11P} and very hard injection functions \citep[$\alpha \leq1.5$;][]{2015MNRAS.448..910C}. In addition, some studies suggest that the magnetic field in the inner jet of blazars is typically lower than 10~G \citep{2009MNRAS.400...26O, 2014JCAP...09..003M}. If the hard-TeV spectrum is still explained by the proton synchrotron emission in a magnetic field $\lesssim10~\rm G$, the maximum proton energy larger than that obtained from the Hillas condition \citep[$E_{\rm Hillas}$;][]{1984ARA&A..22..425H} has to be assumed. The cascade emission from the UHECR can explain the hard-TeV spectrum either, but may also needs extremely high maximum proton energy higher than $E_{\rm Hillas}$ \citep[cf., \citet{2020ApJ...889..149D}]{2016ApJ...817...59T}. Also, the hard-TeV spectra of some BL Lacs show variabilities \citep[e.g.,][]{2010ApJ...709L.163A}, which disfavour the cascade emission from UHECR \citep[e.g.,][]{2012ApJ...757..183P}. Over all, the hard-TeV spectra pose a challenge to conventional one-zone models.

In various one-zone models introduced above, the one-zone  hadronuclear ($pp$) model has not been studied comprehensively. In the study of radiation mechanisms of blazars' jet, the $pp$ interaction is normally neglected, since the particle density in the jet is considered not suffient \citep{2003ApJ...586...79A}. Recently, several associations between high-energy neutrinos and blazars are discovered \citep[e.g.,][]{2018Sci...361.1378I}. However, these events are difficult to be explained by the conventional one-zone $p\gamma$ model \citep[e.g.,][]{2019ApJ...886...23X, 2021ApJ...906...51X}, therefore several innovative $pp$ models are proposed \citep[e.g.,][]{2018ApJ...866..109S, 2019PhRvD..99j3006B, 2019PhRvD..99f3008L, 2022RAA....21..305W}. Also, \cite{2019ApJ...871...81X} suggest that the one-zone $pp$ model may provide a possible solution to reproduce the hard TeV spectrum with a sub-Eddington jet power, since the efficiency of $pp$ interactions is not related to the opacity of $\gamma \gamma$ absorption. Therefore, further considering the $pp$ interaction in the one-zone model comprehensively is a necessary complement to understand the origin of hard-TeV spectra of BL Lacs. 

In this paper, we will analytically study whether the hard-TeV spectra of BL Lacs can be explained without violating basic observations and theories in one-zone $pp$ models. In Section~\ref{M}, we present analytical methods to find the parameter space, and then study if the charge neutrality condition can be satisfied in the framework of one-zone $pp$ models. We present the discussion and the conclusion in Section~\ref{DC}. Throughout the paper, the $\Lambda$CDM cosmological parameters $H_0=70~{\rm km~s^{-1}~Mpc^{-1}}$, $\Omega_{\rm m}=0.3$, $\Omega_{\Lambda}=0.7$ are adopted.

\section{Analytical calculations}\label{M}
For blazar jets, the $pp$ process is usually considered in interactions between the jet and its surrounding materials, such as dense clouds in the BLR, and red giant stars captured from the host galaxy \citep[e.g.,][]{2012A&A...539A..69B}. However, it is unclear if the jet inside has sufficient cold protons. Here, we propose analytical methods to study if the hard-TeV spectra of BL Lacs can be explained by the one-zone $pp$ model, which considering the $pp$ interactions occur in the jet without introducing extreme parameters. We also study if the charge neutrality condition of jet can be satisfied, although this commonly assumed condition \citep{2014Natur.515..376G} is under debate at present \citep{2021ApJ...906..105C}.

It should be noted that the jet composition is currently uncertain. In addition to electrons and protons, there may be a certain number of other charged particles in the jet, such as positrons \citep[e.g.,][]{2016ApJ...831..142M}. In a jet that satisfies the charge neutrality condition, the contribution from $pp$ interactions would becomes weaker if there is a large number of positrons, thus making the model parameters more extreme. In this work, our main purpose is to study the maximum parameter space under the one-zone $pp$ model, therefore we boldly assume that the charged particles in the jet are only relativistic and non-relativistic electrons and protons.

The following analytical calculation of searching the parameter space is under the framework of the conventional one-zone model. Assuming that all the observed jet's non-thermal radiation comes from a single spherical region (hereafter, the blob) composed of a plasma of charged particles in a uniformly entangled magnetic field $B$ with radius $R$ and moving with the bulk Lorentz factor $\Gamma=\frac{1}{\sqrt{1-\beta^2}}$, where $\beta c$ is the speed of the blob, at a viewing angle with respect to observers' line of sight. For the relativistic jet close to the line of sight in blazars with a viewing angle of $\theta \lesssim 1/\Gamma$, we have the Doppler factor $\delta \approx \Gamma$. In this section, the parameters with superscript ``obs'' are measured in the observers' frame, with superscript ``AGN'' are measured in the AGN frame, whereas the parameters without the superscript are measured in the comoving frame, unless specified otherwise.
 
\subsection{Methods}\label{a}
For the hard-TeV spectrum, since the SSC emission cannot explain it because of the KN effect, we suppose that it can be interpreted by the $\gamma$-ray from $\pi^0$ decay produced in the $pp$ interactions. The proton injection luminosity can be calculated as
\begin{equation}\label{L_p}
L_{\rm p,inj}=\frac{4}{3}\pi R^3 \langle \gamma_{\rm p} \rangle m_{\rm p}c^2 \dot{n}_{\rm p}^{\rm inj},
\end{equation}
where $m_{\rm p}$ is the rest mass of a proton, $c$ is the speed of light, 
\begin{equation}
\begin{split}
\langle \gamma_{\rm p} \rangle &= \frac{\int \gamma_{\rm p}\dot{n}_{\rm p}^{\rm inj}(\gamma_{\rm p})d\gamma_{\rm p}}{\dot{n}_{\rm p}^{\rm inj}}\\
&=\frac{(1-\alpha_{\rm p})}{(2-\alpha_{\rm p})}(\frac{\gamma_{\rm p, max}^{2-\alpha_{\rm p}}-\gamma_{\rm p, min}^{2-\alpha_{\rm p}}}{\gamma_{\rm p, max}^{1-\alpha_{\rm p}}-\gamma_{\rm p, min}^{1-\alpha_{\rm p}}})
\end{split}
\end{equation}
represents the average proton Lorentz factor, where $\dot{n}_{\rm p}^{\rm inj} (\gamma_{\rm p})=\dot{n}_{\rm 0,p}\gamma_{\rm p}^{-\alpha_{\rm p}}, \gamma_{\rm p,min}<\gamma_{\rm p}<\gamma_{\rm p,max}$ is the injection rate of proton energy distribution, $\dot{n}_{\rm 0,p}$ is the normalization in units of $\rm cm^{-3}~s^{-1}$, $\alpha_{\rm p}$ is the proton spectral index, $\gamma_{\rm p}$ is the proton Lorentz factor, $\gamma_{\rm p,min}$ is the minimum proton Lorentz factor, and $\gamma_{\rm p,max}$ is the maximum proton Lorentz factor, and $\dot{n}_{\rm p}^{\rm inj}=\int \dot{n}_{\rm p}^{\rm inj}(\gamma_{\rm p})d\gamma_{\rm p}$ is the energy-integrated number density of injected relativistic protons. This $pp$ explanation basically has two constraints, in which one is that the generated $\gamma$-ray luminosity from $\pi^0$ decay exceeds the observed TeV luminosity\footnote{Since the intrinsic hard-TeV spectrum may not be truncated at the maximum energy currently observed, but will extend to higher energy, the theoretical luminosity should be larger than the current observed TeV luminosity.}, i.e.,
\begin{equation}\label{L_TeV}
    L_{\rm TeV}^{\rm obs}\leqslant L_{\rm p,inj}f_{\pi_0}\delta^4,
\end{equation}
where
\begin{equation}\label{f_pi0}
    f_{\pi_0}\approx \frac{1}{3}K_{\rm pp}\sigma_{\rm pp}n_{\rm H}R
\end{equation}
is the $pp$ interaction efficiency, the factor 1/3 is the branching ratio into $\pi_0$, $\sigma_{\rm pp}\approx6\times10^{-26}~{\rm cm^{2}}$ is the cross section for the $pp$ interactions, $n_{\rm H}$ is the number density of cold protons in the jet and $K_{\rm pp}\approx 0.5$ is the inelasticity coefficient \citep{2006PhRvD..74c4018K}. The other constraint is that the total jet power that dominated by the power of relativistic and non-relativistic protons cannot exceed the Eddington luminosity of the SMBH, otherwise the growth of the SMBH would be too quick \citep{2012ApJ...756..116A}. Therefore, we have
\begin{equation}\label{criterion}
L_{\rm H}^{\rm AGN}+L_{\rm p,inj}^{\rm AGN}\leq L_{\rm Edd},
\end{equation}
where $L_{\rm p,inj}^{\rm AGN}=L_{\rm p,inj}\delta^2$ is the the power of the injected relativistic protons in the AGN frame, 
\begin{equation}\label{ledd}
L_{\rm Edd} = 2\pi m_{\rm p}c^3 R_{\rm S} / \sigma_{\rm T}
\end{equation}
is the Eddington luminosity of the SMBH, $\sigma_{\rm T}$ is the Thomson scattering cross section, $R_{\rm S}=2GM_{\rm BH}/c^2$ is the Schwarzschild radius of the SMBH, $M_{\rm BH}$ is the SMBH mass and
\begin{equation}\label{P_H}
    L_{\rm H}^{\rm AGN} = \pi R^2c\delta^2 m_{\rm p}c^2n_{\rm H} 
\end{equation}
is the kinetic power in cold protons. In order to find the maximum parameter space, we assume that the $\pi^0$ decay generates the required minimum $\gamma$-ray luminosity, i.e., $L_{\rm TeV}^{\rm obs}$. Then, substituting $L_{\rm TeV}^{\rm obs}= L_{\rm p,inj}f_{\pi_0}\delta^4$, Eq.~\ref{f_pi0}, Eq.~\ref{ledd} and Eq.~\ref{P_H} into Eq.~\ref{criterion}, we have
\begin{equation}\label{33}
\frac{3\sigma_{\rm T}}{\sigma_{\rm pp}}(\frac{L^{\rm obs}_{\rm TeV}}{L_{\rm Edd}})(\frac{\it{L}_{\rm p,inj}^{\rm AGN}}{L_{\rm Edd}})^{-1}(\frac{R}{R_{\rm S}})+\frac{\it{L}_{\rm p,inj}^{\rm AGN}}{\it{L}_{\rm Edd}}\leqslant  1.
\end{equation}
If $L_{\rm p,inj}^{\rm AGN}$ is regarded as an independent variable, Eq.~\ref{33} becomes a quadratic formula. It can be found that the quadratic formula could get the minimum value when $L_{\rm p,inj}^{\rm AGN}$ is $L_{\rm Edd}/2$. Then substituting $L_{\rm p,inj}^{\rm AGN}=L_{\rm Edd}/2$ into Eq.~\ref{33}, the maximum parameter space of $R$ can be obtained through
\begin{equation}\label{m1}
\frac{R}{R_{\rm S}}\leq \frac{\sigma_{\rm pp}}{12\sigma_{\rm T}}\frac{L_{\rm Edd}}{L_{\rm TeV}^{\rm obs}},
\end{equation}
when the SMBH mass and TeV luminosity are measured.


\begin{table*} 
\caption{The sample of hard-TeV BL Lacs collected by \cite{2019ApJ...871...81X}.}\label{tabel1}
\centering
\begin{tabular}{ccccccc}
\hline\hline
Object & $L_{\rm TeV}^{\rm obs}$ & ${\rm Log}M_{\rm BH}$ & $R_{\rm s}$ & $R$\\
    &  $\rm (erg~s^{-1})$  &   ($\rm M_{\bigodot}$)     & (cm) & (cm) \\
\hline
1ES 0229+200 & $1.00\times10^{45}$ & $9.16\pm0.11$ \citep{2012AA...542A..59M} & $4.29\times10^{14}$ &  $7.00\times10^{14}$ \\
1ES 0347-121 & $1.63\times10^{45}$ & $8.02\pm0.11$ \citep{2012AA...542A..59M} & $3.10\times10^{13}$ &  $2.25\times10^{12}$ \\
1ES 0414+009 & $1.54\times10^{45}$ & $9$ & $2.96 \times10^{14}$ &  $2.17\times10^{14}$ \\
1ES 1101-232 & $1.02\times10^{45}$ & $9$ & $2.96 \times10^{14}$ &  $3.30\times10^{14}$ \\
1ES 1215+303 & $2.11\times10^{45}$ & $8.4$ \citep{2012NewA...17....8G} & $7.45\times10^{13}$ &  $1.00\times10^{13}$ \\
1ES 1218+304 & $5.45\times10^{45}$ & $8.04\pm0.24$ \citep{2012AA...542A..59M} & $3.25\times10^{13}$  & $7.40\times10^{11}$ \\
S5 0716+714  & $3.33\times10^{46}$ & $8$ \citep{2015MNRAS.450L..21Z} & $2.96\times10^{13}$ &  $1.01\times10^{11}$ \\
W Comae   & $6.00\times10^{44}$ & $8.7$ \citep{2015MNRAS.450L..21Z} & $1.49\times10^{14}$ &  $1.41\times10^{14}$ \\
TXS 0506+056 & $1.74\times10^{45}$ & $9$ & $2.96\times10^{14}$ &  $1.92\times10^{14}$ \\
\hline
\label{table1}
\end{tabular}

\textcolor{blue}{Notes.} Columns from left to right: the source name; the observed luminosity of the TeV spectrum; logarithm of the SMBH mass in units of the solar mass, $\rm M_{\odot}$; the Schwarzschild radius of the SMBH; the maximum $R$ obtained by Eq.~\ref{m1}.
\end{table*}

In the above analytical calculations, by fixing the $\gamma$-ray luminosity from $\pi^0$ decay to the required minimum value, and keeping the total jet power not exceeding the Eddington luminosity of SMBH, the parameter space of $R$ is well constrained. Similarly, the maximum parameter space of $R$ can also be obtained by fixing the total jet power to the Eddington luminosity of SMBH, which is the maximum value that can be set, and making the $\gamma$-ray luminosity from $\pi^0$ decay larger than the observed TeV luminosity. Here, the power of the injected relativistic protons is considered as a fraction $\chi_{\rm p}$ of the total jet power, which is 
\begin{equation}\label{L_p,inj}
    L_{\rm p,inj}^{\rm AGN}=\chi_{\rm p}L_{\rm Edd}.
\end{equation}
Since the power of the relativistic and non-relativistic electrons is normally negligible compared to that of relativistic and non-relativistic protons, according to Eq.~\ref{P_H} and Eq.~\ref{L_p,inj}, $n_{\rm H}$ can be estimated as
\begin{equation}\label{n_H'}
    n_{\rm H}=\frac{(1-\chi_{\rm p})L_{\rm Edd}}{\pi R^2\delta^2m_{\rm p}c^3}.
\end{equation}
The $\gamma$-ray luminosity from $\pi_0$ decay produced in the $pp$ interactions can be given by
\begin{equation}\label{L_gamma}
    L_{\gamma,{\rm \pi_0}}^{\rm AGN}\approx L_{\rm p,inj}^{\rm AGN} f_{\pi_0}.
\end{equation}
Combining with Eq.~\ref{f_pi0}, Eq.~\ref{L_p,inj}, Eq.~\ref{n_H'} and Eq.~\ref{L_gamma}, if the hard-TeV spectrum can be interpreted by the $\gamma$-ray from $\pi_0$ decay produced in the $pp$ interactions, we can get
\begin{equation}\label{L_gamma obs}
     L_{\rm TeV}^{\rm obs}\leqslant L_{\gamma,{\rm \pi_0}}^{\rm obs}=\frac{\sigma_{\rm pp}\chi_{\rm p}(1-\chi_{\rm p})L_{\rm Edd}^2}{6\pi m_{\rm p}c^3R}.
\end{equation}
In order to get the maximum parameter space, it is necessary to set $\chi_{\rm p}=0.5$, which is consistent with the result obtained below Eq.~\ref{33}. If further substituting Eq.~\ref{ledd} into Eq.~\ref{L_gamma obs}, Eq.~\ref{m1} can also be derived.

\cite{2019ApJ...871...81X} collected a sample of hard-TeV BL Lacs with measured TeV luminosities and the SMBH masses. Their corresponding maximum $R$ obtained by Eq.~\ref{m1} are shown in Table~\ref{table1}. It can be seen that the derived maximum $R$ is quite small which is comparable or even smaller than the Schwarzschild radius of the SMBH $R_{\rm S}$. However, in such a compact blob, TeV photons are likely to be absorbed due to the internal $\gamma \gamma$ absorption. Since the spectra shape of soft photons is fixed by observational data points, i.e., the low energy component, the internal $\gamma \gamma$ absorption optical depth is only related to $R$ and $\delta$, i.e., $\tau_{\gamma\gamma} \propto R^{-1}\delta^{-4}$. \cite{2019ApJ...871...81X} give the value of $R$ and $\delta$ when $\tau_{\gamma\gamma}$ of the maximum energy of the hard-TeV spectrum is equal to unity. Therefore, in order to prevent the hard-TeV spectrum being absorbed, a larger $\delta$ must be introduced, since the maximum $R$ shown in Table~\ref{table1} is already much smaller than that given by \cite{2019ApJ...871...81X}. If the maximum $\delta$ that can be set is 30 as indicated by observation \citep{2009A&A...494..527H}, only 1ES 0229+200 has a certain parameter space which can be used to interpret the hard-TeV spectrum with the one-zone $pp$ model.

\subsection{Charge neutrality condition of jet}
Recently, \cite{2019PhRvD..99j3006B} suggested that when considering the charge neutrality condition of blazar jet, there will be sufficient cold protons, making the $pp$ interaction to be efficient and explaining the observed TeV $\gamma$-ray and neutrino emission from the blazar TXS 0506+056. However, if the kinetic power of cold protons is taken into account, a super-Eddington jet power is still needed since the blob radius introduced in their modeling is about a hundred times larger than that obtained by our analytical methods. In the following, we will investigate if the charge neutrality condition of jet can be satisfied when interpreting the hard-TeV spectra without introducing a super-Eddington jet power.

Due to the lack of the strong emission from external photon fields in TeV BL Lacs, it is generally accepted that the emission below $\sim100$ GeV is originated from the synchrotron and SSC emission from primary relativistic electrons. The electron injection luminosity can be calculated as
\begin{equation}\label{L_e}
L_{\rm e,inj}=\frac{4}{3}\pi R^3 \langle \gamma_{\rm e}\rangle m_{\rm e}c^2 \dot{n}_{\rm e}^{\rm inj},
\end{equation}
where $m_{\rm e}$ is the rest mass of an electron, 
\begin{equation}
\begin{split}
\langle \gamma_{\rm e} \rangle &= \frac{\int \gamma_{\rm e}\dot{n}_{\rm e}^{\rm inj}(\gamma_{\rm e})d\gamma_{\rm e}}{\dot{n}_{\rm e}^{\rm inj}}\\
&=\frac{(1-\alpha_{\rm e})}{(2-\alpha_{\rm e})}(\frac{\gamma_{\rm e, max}^{2-\alpha_{\rm e}}-\gamma_{\rm e, min}^{2-\alpha_{\rm e}}}{\gamma_{\rm e, max}^{1-\alpha_{\rm e}}-\gamma_{\rm e, min}^{1-\alpha_{\rm e}}})
\end{split}
\end{equation}
is the average electron Lorentz factor, where $\dot{n}_{\rm e}^{\rm inj}(\gamma_{\rm e})=\dot{n}_{\rm 0,e}\gamma_{\rm e}^{-\alpha_{\rm e}}, \gamma_{\rm e,min}<\gamma_{\rm e}<\gamma_{\rm e,max}$ is the injection rate of electron energy distribution (EED), $\dot{n}_{\rm 0,e}$ is the normalization in units of $\rm cm^{-3}~s^{-1}$, $\alpha_{\rm e}$ is the electron spectral index, $\gamma_{\rm e}$ is the electron Lorentz factor, $\gamma_{\rm e,min}$ is the minimum electron Lorentz factor, and $\gamma_{\rm e,max}$ is the maximum electron Lorentz factor, and $\dot{n}_{\rm e}^{\rm inj}=\int \dot{n}_{\rm e}^{\rm inj}(\gamma_{\rm e})d\gamma_{\rm e}$ is the energy-integrated number density of injected relativistic electrons.
Taking into account cold electrons that are not accelerated or cooled, the total number density of electrons in the jet $n_{\rm e,tot}$ can be approximated as 
\begin{equation}\label{n_e,tot}
n_{\rm e,tot}=\chi_{\rm e} n_{\rm e}, 
\end{equation}
where $\chi_{\rm e}$ represents the ratio of $n_{\rm e,tot}$ to $n_{\rm e}$ and
\begin{equation}
 n_{\rm e}\approx t_{\rm dyn}\dot{n}_{\rm e}^{\rm inj}
 \end{equation}
 is the number density of relativistic electrons, where $t_{\rm dyn}=R/c$ is the dynamical timescale of the blob\footnote{This relation comes from the fact that the radiative cooling of most electrons is in the slow cooling regime for BL Lacs. Although the cooling of high-energy electrons might be in the fast cooling regime, this result will not be significantly changed because the number density of relativistic electrons $n_{\rm e,rel}$ is dominated by the low-energy electrons that dissipated in the slow cooling regime.}.
 
\begin{figure}[htbp]
\centering
\includegraphics[width=1.1\columnwidth]{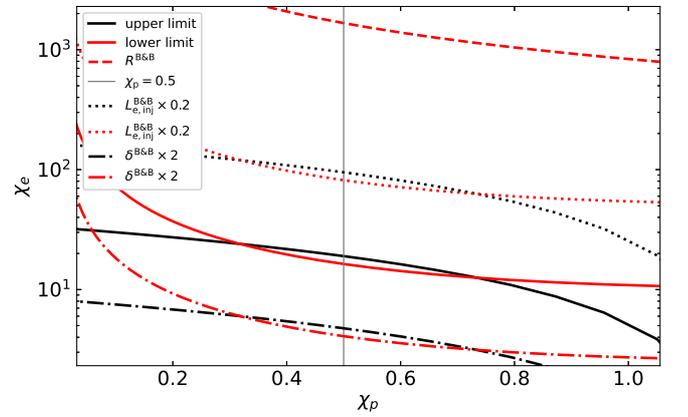}
\caption{The comparison of $\chi_{\rm e}$ as a function of $\chi_{\rm p}$ derived by Eq.~\ref{ineq} with different parameters. For all the line styles, the black and red curves represent the upper and lower limits, respectively. The solid curves are obtained when setting $R$ as the maximum value given in Table~\ref{table1} and values of other parameters as those provided by \cite{2019PhRvD..99j3006B}. Based on this, the dotted curves show the upper and lower limits when setting the electron injection luminosity as $L_{\rm e, inj}^{\rm B\&B}\times 0.2$, the dashed-dotted curves show the upper and lower limits when setting the Doppler factor as $\delta^{\rm B\&B}\times 2$, and the dashed curve shows the lower limit when setting the blob radius as used in \cite{2019PhRvD..99j3006B}. The vertical line represents $\chi_{\rm p}=0.5$. \label{chiep}}
\end{figure}

Since the cooling of relativistic protons in the jet is inefficient, the number density of relativistic protons can be approximated as
\begin{equation}\label{n_p,rel}
    n_{\rm p}\approx t_{\rm dyn}\dot{n}_{\rm p}^{\rm inj}.
\end{equation}
Under the charge neutrality condition, combining Eq.~\ref{n_e,tot} and Eq.~\ref{n_p,rel}, we get the number density of cold protons in the jet through
\begin{equation}\label{n_H}
    n_{\rm H}=n_{\rm e,tot}-n_{\rm p}.
\end{equation}
Substituting Eq.~\ref{L_p} and Eq.~\ref{L_e} into Eq.~\ref{n_H}, we have
\begin{equation}\label{nhchi_w}
\begin{split}
n_{\rm H}=t_{\rm dyn}(\frac{3\chi_{\rm e}L_{\rm e, inj}}{4\pi R^3\langle \gamma_{\rm e} \rangle m_{\rm e}c^2}-\frac{3L_{\rm p, inj}}{4\pi R^3\langle \gamma_{\rm p} \rangle m_{\rm p}c^2}).
\end{split}
\end{equation}
Further substituting Eq.~\ref{nhchi_w} into Eq.~\ref{L_TeV} and Eq.~\ref{criterion}, we can get
\begin{equation}\label{ineq}
\frac{\langle \gamma_{\rm e} \rangle}{L_{\rm e,inj}}\frac{m_{\rm e}}{m_{\rm p}}(4\frac{\sigma_{\rm T}}{\sigma_{\rm pp}}\frac{R}{R_{\rm S}}\frac{L_{\rm TeV}^{\rm obs}}{\chi_{\rm p}\delta^2}+\frac{L_{\rm p,inj}}{\langle \gamma_{\rm p} \rangle}) \leq \chi_{\rm e} \leq \langle \gamma_{\rm e} \rangle \frac{L_{\rm p,inj}}{L_{\rm e,inj}}\frac{m_{\rm e}}{m_{\rm p}}\left[\frac{4(1-\chi_p)}{3\chi_p} + \frac{1}{\langle \gamma_{\rm p} \rangle}\right].
\end{equation}
With the above inequalities, the range of $\chi_{\rm e}$ satisfying the charge neutrality condition can be obtained. It is difficult to give a concise expression of the above inequalities, since the number density of relativistic electrons is greatly affected by the shape and energy range of EED, and $\gamma_{\rm e, min}$ and $\gamma_{\rm e, max}$ are poorly constrained in one-zone models. In any case, the above inequalities suggest that even though $R$ is small, $\chi_{\rm e}$ still has a parameter space to satisfy the charge neutrality condition. Here we take TXS 0506+056 as an example. If we set $R$ as its maximum value given in Table~\ref{table1}, and values of other parameters as those provided by \cite{2019PhRvD..99j3006B}\footnote{i.e., $\delta=10, L_{\rm e, inj}=2.3\times10^{42}~\rm erg/s, \gamma_{\rm e,min}=1, \gamma_{\rm e,max}=1.5\times10^5, \alpha_{\rm e}=1.71, \gamma_{\rm p,min}=1, \gamma_{\rm p,max}=1.1\times10^7, \alpha_{\rm p}=2.13$.}, the range of $\chi_{\rm e}$ as a function of $\chi_{\rm p}$ that obtained by Eq.~\ref{ineq} is shown by the intersection area of black (upper limit) and red (lower limit) solid curves in Fig.~\ref{chiep}. The vertical line in Fig.~\ref{chiep} marks the position of $\chi_{\rm p}=0.5$. It can be found that when $\chi_{\rm p}=0.5$, $\chi_{\rm e}$ has largest range of value, which is consistent with the result obtained in Section \ref{a}. In addition, the comparison results of adjusting some other parameters are also shown in Fig.~\ref{chiep}. It can be seen that adjusting the electron injection luminosity (dotted curves) and the Doppler factor (dashed-dotted curves) will change the parameter space of $\chi_{\rm e}$ but the size of intersection area of the upper and lower limits remains the same. Also, when setting a larger blob radius $R^{\rm B\&B}=2.2\times10^{16}~\rm cm$ as used in \cite{2019PhRvD..99j3006B}, the lower limit represented by red dashed curve is much higher than the upper limit represented by black solid curve (since adjusting blob radius do not affect the upper limit, the black dashed curve is overlapped by the black solid curve), which suggest that a highly super-Eddington jet power has to be introduced.

\section{Discussion and conclusion}\label{DC}
In this work, we have investigated if one-zone $pp$ models can explain the hard-TeV spectra of BL Lacs. With analytical calculations, we find that an extremely compact radiation region has to be assumed, if the hard-TeV spectrum is interpreted by $\gamma$ rays produced from $\pi_0$ decay in $pp$ interaction. However, if further considering the internal $\gamma \gamma$ opacity at TeV band is less than unity and the Doppler factor should be smaller than 30, only the hard-TeV spectra of 1ES 0229+200 can be explained. Even though, the allowable maximum radius of the blob under the constraints of Eq.~\ref{m1} is very small, comparable to the Schwarzschild radius of the SMBH. This compact blob might be a relativistically moving plasmoid generated in the magnetic reconnection of the jet inside \citep{2017ApJ...841...61A} and imply a fast minute-scale variability, which is observed in TeV band of radio galaxies \citep[e.g.,][]{2014Sci...346.1080A} and blazars \citep[e.g.,][]{2007ApJ...669..862A}. However, no evidence of fast variability in TeV band of 1ES 0229+200 is found, therefore the injection of relativistic protons in such a compact blob must be continuous. On the other hand, it can be seen from Eq.~\ref{m1} that for AGNs with low TeV luminosity, such as radio galaxies, the effective $pp$ processes can occur in a large scale region (e.g., lobes) far from the SMBH, which could be a possible origin of the TeV spectrum \citep[e.g.,][]{2016A&A...595A..29S}. 

In addition to conventional radiation models of blazars, another possible solution that invoked hypothetical conversion between axion-like particles (ALPs) and propagating $\gamma$-ray photons in an external magnetic field has been widely investigated on explaining the TeV emission of blazars \citep[e.g.,][]{2009PhRvD..79l3511S}. The conversion could enable propagating TeV-photons to avoid $\gamma \gamma$ absorption and thus lead to a very hard-TeV spectrum as observed \citep[e.g.,][]{2020PhRvD.101f3004L, 2021PhRvD.104h3014L}. If it takes place in the blob of the inner jet with a strong magnetic field, the effective conversion occurs on the order of GeV and the ALPs converted from the source photons cannot reconvert into photons in the lower Galactic magnetic-field \citep{2015PhLB..744..375T}. As a result, a substantial part of GeV photons would be lost, and it may lead to the exceeded Eddington luminosity of the SMBH as well. Conversely, if it occurs in the large-scale jet (also the TeV emission region) with a weaker magnetic field, this problem would be avoided. 

In fact, the discovered different variability patterns between TeV emission and other wavelength emission of some hard-TeV BL Lacs may support the multi-zone origin of the hard-TeV spectra. For example, a TeV flare of the hard-TeV BL Lac 1ES 1215+303 was found by MAGIC in 2011, while no variability was found in the GeV band at the same time \citep{2012A&A...544A.142A}. Also, no fast variability in TeV band is found for many hard-TeV BL Lacs, while other wavelength emissions are highly variable \citep[e.g., 1ES 0229+200;][]{2014ApJ...782...13A}.

To summarize, we investigate the possibility that if the hard-TeV spectra of BL Lacs can be explained with the one-zone $pp$ model, which complemented the completeness of the one-zone model in the study of the hard-TeV spectra of BL Lacs. Unfortunately, only 1ES 0229+200 can be explained at the cost of introducing a very small blob radius that is comparable to the Schwarzschild radius of SMBH. Therefore, combined with previous studies that applied other one-zone models, we suggest that any one-zone model cannot explain the hard-TeV spectra of BL Lacs generally. It should originate from multiple emitting regions \citep[e.g.,][]{2012MNRAS.424.2173Y}. The region generating the emission below $\sim100~\rm GeV$ and the region producing the hard-TeV spectrum should be decoupled \citep{2022PhRvD.105b3005W}. Both leptonic and hadronic processes are possible to explain the hard-TeV spectrum of BL Lacs.

\begin{acknowledgements}
This work was supported by JSPS KAKENHI Grant JP19H00693. This work was supported in part by a RIKEN pioneering project "Evolution of Matter in the Universe (r-EMU)".
\end{acknowledgements}



\begin{thebibliography}{99}
\bibitem[Acciari et al.(2010)]{2010ApJ...709L.163A} Acciari, V.~A., Aliu, E., Beilicke, M., et al.\ 2010, \apjl, 709, L163
\bibitem[Ackermann et al.(2016)]{2016PhRvL.116o1105A} Ackermann, M., Ajello, M., Albert, A., et al.\ 2016, \prl, 116, 151105
\bibitem[Aharonian(2000)]{2000NewA....5..377A} Aharonian, F.~A.\ 2000, \na, 5, 377
\bibitem[Aharonian et al.(2017)]{2017ApJ...841...61A} Aharonian, F.~A., Barkov, M.~V., \& Khangulyan, D.\ 2017, \apj, 841, 61
\bibitem[Albert et al.(2007)]{2007ApJ...669..862A} Albert, J., Aliu, E., Anderhub, H., et al.\ 2007, \apj, 669, 862
\bibitem[Aleksi{\'c} et al.(2012)]{2012A&A...544A.142A} Aleksi{\'c}, J., Alvarez, E.~A., Antonelli, L.~A., et al.\ 2012, \aap, 544, A142
\bibitem[Aleksi{\'c} et al.(2014)]{2014Sci...346.1080A} Aleksi{\'c}, J., Ansoldi, S., Antonelli, L.~A., et al.\ 2014, Science, 346, 1080
\bibitem[Aliu et al.(2014)]{2014ApJ...782...13A} Aliu, E., Archambault, S., Arlen, T., et al.\ 2014, \apj, 782, 13
\bibitem[Antognini et al.(2012)]{2012ApJ...756..116A} Antognini, J., Bird, J., \& Martini, P.\ 2012, \apj, 756, 116
\bibitem[Atoyan \& Dermer(2003)]{2003ApJ...586...79A} Atoyan, A.~M. \& Dermer, C.~D.\ 2003, \apj, 586, 79
\bibitem[Banik \& Bhadra(2019)]{2019PhRvD..99j3006B} Banik, P. \& Bhadra, A.\ 2019, \prd, 99, 103006
\bibitem[Bosch-Ramon et al.(2012)]{2012A&A...539A..69B} Bosch-Ramon, V., Perucho, M., \& Barkov, M.~V.\ 2012, \aap, 539, A69
\bibitem[B{\"o}ttcher et al.(2013)]{2013ApJ...768...54B} B{\"o}ttcher, M., Reimer, A., Sweeney, K., et al.\ 2013, \apj, 768, 54
\bibitem[B{\l}a{\.z}ejowski et al.(2000)]{2000ApJ...545..107B} B{\l}a{\.z}ejowski, M., Sikora, M., Moderski, R., et al.\ 2000, \apj, 545, 107
\bibitem[Cerruti et al.(2015)]{2015MNRAS.448..910C} Cerruti, M., Zech, A., Boisson, C., et al.\ 2015, \mnras, 448, 910
\bibitem[Chen \& Zhang(2021)]{2021ApJ...906..105C} Chen, L. \& Zhang, B.\ 2021, \apj, 906, 105
\bibitem[Das et al.(2020)]{2020ApJ...889..149D} Das, S., Gupta, N., \& Razzaque, S.\ 2020, \apj, 889, 149
\bibitem[Deng et al.(2021)]{2021MNRAS.506.5764D} Deng, X.-J., Xue, R., Wang, Z.-R., et al.\ 2021, \mnras, 506, 5764
\bibitem[Dermer \& Schlickeiser(1993)]{1993ApJ...416..458D} Dermer, C.~D. \& Schlickeiser, R.\ 1993, \apj, 416, 458
\bibitem[Essey et al.(2011)]{2011APh....35..135E} Essey, W., Ando, S., \& Kusenko, A.\ 2011, Astroparticle Physics, 35, 135
\bibitem[Ghisellini et al.(2014)]{2014Natur.515..376G} Ghisellini, G., Tavecchio, F., Maraschi, L., et al.\ 2014, \nat, 515, 376
\bibitem[Gupta et al.(2012)]{2012NewA...17....8G} Gupta, S.~P., Pandey, U.~S., Singh, K., et al.\ 2012, \na, 17, 8
\bibitem[Hillas(1984)]{1984ARA&A..22..425H} Hillas, A.~M.\ 1984, \araa, 22, 425
\bibitem[Hovatta et al.(2009)]{2009A&A...494..527H} Hovatta, T., Valtaoja, E., Tornikoski, M., et al.\ 2009, \aap, 494, 527
\bibitem[IceCube Collaboration et al.(2018)]{2018Sci...361.1378I} IceCube Collaboration, Aartsen, M.~G., Ackermann, M., et al.\ 2018, Science, 361, eaat1378
\bibitem[Katarzy{\'n}ski et al.(2006)]{2006MNRAS.368L..52K} Katarzy{\'n}ski, K., Ghisellini, G., Tavecchio, F., et al.\ 2006, \mnras, 368, L52
\bibitem[Kelner et al.(2006)]{2006PhRvD..74c4018K} Kelner, S.~R., Aharonian, F.~A., \& Bugayov, V.~V.\ 2006, \prd, 74, 034018
\bibitem[Liu et al.(2019)]{2019PhRvD..99f3008L} Liu, R.-Y., Wang, K., Xue, R., et al.\ 2019, \prd, 99, 063008
\bibitem[Long et al.(2020)]{2020PhRvD.101f3004L} Long, G.~B., Lin, W.~P., Tam, P.~H.~T., et al.\ 2020, \prd, 101, 063004
\bibitem[Long et al.(2021)]{2021PhRvD.104h3014L} Long, G., Chen, S., Xu, S., et al.\ 2021, \prd, 104, 083014
\bibitem[Madejski et al.(2016)]{2016ApJ...831..142M} Madejski, G.~M., Nalewajko, K., Madsen, K.~K., et al.\ 2016, \apj, 831, 142
\bibitem[Marscher \& Gear(1985)]{1985ApJ...298..114M} Marscher, A.~P. \& Gear, W.~K.\ 1985, \apj, 298, 114
\bibitem[Meyer et al.(2012)]{2012AA...542A..59M} Meyer, M., Raue, M., Mazin, D., et al.\ 2012, \aap, 542, A59
\bibitem[Meyer et al.(2014)]{2014JCAP...09..003M} Meyer, M., Montanino, D., \& Conrad, J.\ 2014, \jcap, 2014, 003
\bibitem[O'Sullivan \& Gabuzda(2009)]{2009MNRAS.400...26O} O'Sullivan, S.~P. \& Gabuzda, D.~C.\ 2009, \mnras, 400, 26
\bibitem[Petropoulou \& Dermer(2016)]{2016ApJ...825L..11P} Petropoulou, M. \& Dermer, C.~D.\ 2016, \apjl, 825, L11
\bibitem[Prosekin et al.(2012)]{2012ApJ...757..183P} Prosekin, A., Essey, W., Kusenko, A., et al.\ 2012, \apj, 757, 183
\bibitem[Sahu et al.(2019)]{2019ApJ...884L..17S} Sahu, S., L{\'o}pez Fort{\'\i}n, C.~E., \& Nagataki, S.\ 2019, \apjl, 884, L17
\bibitem[Sahakyan(2018)]{2018ApJ...866..109S} Sahakyan, N.\ 2018, \apj, 866, 109
\bibitem[S{\'a}nchez-Conde et al.(2009)]{2009PhRvD..79l3511S} S{\'a}nchez-Conde, M.~A., Paneque, D., Bloom, E., et al.\ 2009, \prd, 79, 123511
\bibitem[Sikora et al.(1994)]{1994ApJ...421..153S} Sikora, M., Begelman, M.~C., \& Rees, M.~J.\ 1994, \apj, 421, 153
\bibitem[Sun et al.(2016)]{2016A&A...595A..29S} Sun, X.-. na ., Yang, R.-. zhi ., Mckinley, B., et al.\ 2016, \aap, 595, A29
\bibitem[Tavecchio et al.(2015)]{2015PhLB..744..375T} Tavecchio, F., Roncadelli, M., \& Galanti, G.\ 2015, Physics Letters B, 744, 375
\bibitem[Takami et al.(2016)]{2016ApJ...817...59T} Takami, H., Murase, K., \& Dermer, C.~D.\ 2016, \apj, 817, 59
\bibitem[Tan et al.(2020)]{2020ApJS..248...27T} Tan, C., Xue, R., Du, L.-M., et al.\ 2020, \apjs, 248, 27
\bibitem[Urry \& Padovani(1995)]{1995PASP..107..803U} Urry, C.~M. \& Padovani, P.\ 1995, \pasp, 107, 803
\bibitem[Wang \& Xue(2022a)]{2022RAA....21..305W} Wang, Z.-R. \& Xue, R.\ 2022, Research in Astronomy and Astrophysics, 21, 305
\bibitem[Wang et al.(2022b)]{2022PhRvD.105b3005W} Wang, Z.-R., Liu, R.-Y., Petropoulou, M., et al.\ 2022, \prd, 105, 023005
\bibitem[Xue et al.(2019a)]{2019ApJ...871...81X} Xue, R., Liu, R.-Y., Wang, X.-Y., et al.\ 2019, \apj, 871, 81
\bibitem[Xue et al.(2019b)]{2019ApJ...886...23X} Xue, R., Liu, R.-Y., Petropoulou, M., et al.\ 2019, \apj, 886, 23
\bibitem[Xue et al.(2021)]{2021ApJ...906...51X} Xue, R., Liu, R.-Y., Wang, Z.-R., et al.\ 2021, \apj, 906, 51
\bibitem[Yan et al.(2012)]{2012MNRAS.424.2173Y} Yan, D., Zeng, H., \& Zhang, L.\ 2012, \mnras, 424, 2173
\bibitem[Zdziarski \& Bottcher(2015)]{2015MNRAS.450L..21Z} Zdziarski, A.~A. \& Bottcher, M.\ 2015, \mnras, 450, L21
\end{thebibliography}
\end{document}